# LID Framework

# A new method for geospatial and exploratory data analysis of potential innovation determinants at the neighborhood level

**Eleni Oikonomaki** [1,*], **Belivanis Dimitris**[2], **Kakderi Christina**[3]

[1] URENIO Research, Aristotle University of Thessaloniki, 541 24, Thessaloniki, Greece, UC Berkeley, College of Environmental Design, 230 Bauer Wurster Hall, Berkeley, CA 94720-1820; *Corresponding Author: elenoikonomaki@plandevel.auth.gr, elenoikonomaki@gmail.com (E.O.), https://orcid.org/0000-0001-6534-7695

[2] dbelivanis@stanford.edu (D.B), https://orcid.org/0009-0003-2831-8686

[3] URENIO Research, Aristotle University of Thessaloniki, 541 24, Thessaloniki, Greece; kakderi@plandevel.auth.gr (C.K), https://orcid.org/0000-0001-6499-3919

## ABSTRACT

The geography of innovation provides a framework for understanding how territorial characteristics shape the generation of innovation, often through the lenses of spatial and cognitive proximity. Within this literature, there is a notable empirical bias toward national and regional units, with urban and sub-regional geographies receiving comparatively less analytical attention. Existing studies at the local level typically focus on a limited set of indicators (e.g., firm-level data or patents and basic socioeconomic characteristics), with few offering a systematic, replicable framework that integrates urban form, mobility, amenities, and human-capital proxies at the neighborhood scale. Our study conducts an empirical investigation of innovation, focusing on a finer spatial scale than the periphery or city boundaries, and exploring metrics beyond proprietary data or static indicators. It builds the Local Innovation Determinants (LID) database and methodological framework to uncover key enabling factors of innovation across diverse regions, combining well-established and alternative (newly introduced) territorial determinants, complementing traditional data from governmental agencies with publicly available data through modern APIs to obtain a more granular and broader understanding of the spatial dynamics that shape innovation capacity at a finer spatial scale. It uses exploratory big and geospatial data analytics and random forest models across neighborhoods in New York and Massachusetts to identify significant variables across four dimensions: social factors, economic characteristics, land use and urban mobility, morphology, and environment. Our findings indicate that alternative data sources have significant yet underexplored potential to enhance our understanding of innovation dynamics. City policymakers should carefully consider key determinants and distinctive characteristics of individual neighborhoods when formulating and implementing local innovation policies.

**Highlights:**

- Develops a spatial modelling framework and a new methodology to assess technological innovation at lower than regional and urban scales to obtain a more granular and broader understanding of its determinants.

- Constructs a Local Innovation Determinants (LID) database by combining well-established determinants with alternative innovation-enabling factors to conduct a comparative neighborhood analysis, complementing traditional data from governmental agencies with open-source, API-mined data.
- Offers a scalable urban analytics framework and a data pipeline for mining open-source urban data and applies machine learning (feature importance) algorithms to illustrate significant variables across four distinct dimensions: social factors, economic characteristics, land use and urban mobility, environment, and urban morphology.

1. **Introduction: Evolving Geographies of Innovation, Reconsidered Locally**

Cities have historically served as hubs of innovation, benefiting from the agglomeration of talent, knowledge, and entrepreneurial activity. The spatial dimension is fundamental to innovation, as firms and industries tend to cluster in places that offer dense networks of specialized resources, highly skilled human capital, and infrastructure (Porter, 1998; Storper, 1997; Audretsch and Feldman, 1996). This geographic concentration facilitates learning, collaboration, and technological advancement, reinforcing the innovation capacity of these areas over time. The literature in innovation studies has well-documented the importance of environments and spatial clusters in driving innovation and productivity (Davis and Dingel 2019; Arzaghi and Henderson 2008).

To capture innovation outcomes across diverse clusters or regions, empirical studies focus on patent filing, high-growth entrepreneurship backed by venture capital, and/or employment in industries or occupations heavily involved in research and development. Another promising avenue of research is to measure the extent and nature of inventions as they vary across a metropolitan area at a larger scale. These approaches aid in achieving a sufficient sample size, as exemplified by Acs, Anselin, and Varga (2002), Delgado, Porter, and Stern (2010), Glaeser et al. (2009), and G. Carlino and Kerr (2015). While informative, they focus on a limited set of indicators (e.g., firm data, patents, and basic socioeconomics) rather than offering a systematic, replicable framework that integrates urban form, mobility, amenities, and human-capital proxies at the neighbourhood scale.

When selecting a geographical unit, first empirical investigations in the field considered the complete array of nations, states, regions, metropolitan areas, or cities, or within firm and large corporations geographies and individual university buildings, or publicly funded institutions, (Catalini 2018; Roche et al. 2022; Allen 1977; Wallsten 2001), offering little insight into the innovation capacities of their localities. This is inconsistent with the importance of distance for innovation, which predicts that innovation would occur primarily within narrower geographic areas. For example, Kerr and Kominers (2015) and Arzaghi and Henderson (2008) document how knowledge spillovers quickly dissipate when travel time exceeds approximately 15 minutes. More recently, a growing body of research has begun to highlight the relevance of spatial factors, such as walkability, urban compactness, access to amenities, and spatial connectivity, as key components of urban innovation ecosystems (Catalini 2018; Hamidi and Zandiatashbar 2018; Rammer et al., 2019; Roche 2020; Matsiuk et al., 2024). This recent work examines the micro-geography of innovation within cities – from street-level patent clustering and firm location patterns to neighbourhood diversity and innovation districts - to understand the factors that shape not only where innovation occurs but also how it evolves.

To capitalize on this potential, neighbourhood data can be used to capture the granular effects of territorial characteristics, and these insights can be fed into broader policy frameworks. However, the development of such databases poses significant technical and institutional challenges, as neighbourhood data are often scarce, fragmented, or poorly integrated into decision-making processes. Effective monitoring requires a well-orchestrated and robust system architecture capable of aggregating multi-source data. It also requires a holistic approach to the growing adoption of digital technologies, including AI, geospatial, big data analytics, and mapping tools, which can enable more transparent, inclusive, and real-time monitoring of neighbourhood characteristics. Nevertheless, designing a neighbourhood-scale database and methodological framework for innovation assessment that is technically robust is far from straightforward.

Our paper addresses these identified challenges by exploring the innovation generation more systematically at the local level, capturing data beyond commonly examined regional or national units, and at scales larger than the firm or institution level. Our framework builds upon the identified need to incorporate new data sources and indicators that reflect the multifaceted and dynamic nature of innovation, including urban morphology, environmental quality, social infrastructure, mobility, amenities, and socioeconomic variables, into a database and conceptual framework. The intersection of urban form and innovation generation is a key area of inquiry for economists and urban planners. Understanding how the built environment, land use, and mobility infrastructure support knowledge production requires interdisciplinary approaches and diverse neighborhood variables, offering new pathways for policy design and urban development.

In this direction, the structure of the paper proceeds as follows. Section 1 reviews prior studies on the effects of geography on innovation performance, emphasizing the need to understand *innovation systems as groups of places.* Section 2 presents the aims and empirical strategy for combining established and alternative data, creating a database, and developing a methodological framework for assessing local innovation-enabling factors. Section 3 provides an overview of traditional determinants of innovation and introduces new neighborhood variables that may enable innovation performance. Section 4 examines how machine learning models can be used to assess the significance of selected variables in predicting innovation activity at the zip code level. Section 5 concludes with recommendations for the study's next steps to inform innovation policymaking.

## 2. Building the Conceptual Framework for Understanding Innovation Systems Through Place

In recent decades, the understanding of innovation has expanded beyond the traditional model of "R&D in firms and universities" to a broader, more distributed view of user-innovation communities (von Hippel, 2005), where users themselves often develop economically significant products and solutions, many of which are openly shared. This shifts innovation from being solely producer-centric - confined to internal R&D labs - to also being user-centric, emerging from communities, lead users, and collaborative networks. Richard Florida (2012) further connected the presence of knowledge workers, professionals, and creative practitioners to local innovation dynamics. Although aspects of his theories have been debated (Peck, 2005; Glaeser, 2005; Markusen, 2006; Scott, 2006) the central insight is widely acknowledged: places with higher concentrations of knowledge workers tend to show greater innovative capacity, and contextual factors such as cultural amenities, diversity, tolerance, and vibrant public spaces influence where these workers choose to live and work. Therefore, innovation systems are strongly place-based and labor-based, not only R&D expenditure-based.

Innovation spaces such as coworking spaces provide connections, knowledge spillovers, support, and legitimacy, especially for founders, women, minorities, and foreign entrepreneurs (Miller and McAdam, 2015; Kloosterman and Rath, 2001). Also, living labs and other user- and citizen-oriented spaces in real-life settings play a role in the co-creation of innovation (von Hippel, 2005). The main focus of this literature is that a) innovation is increasingly distributed between firms, users, communities, knowledge workers, and hybrid spaces, and that b) the local environment (spaces, amenities, networks, governance) matters for enabling both formal R&D and informal/user innovation. Thus, innovation studies are shifting from viewing the innovation system primarily as a set of institutions (as in NIS/RIS) to understanding it increasingly as a place where spatial proximity, local context, and everyday interactions shape innovation activity.

The vast majority of innovative case studies emphasize the critical role of specific neighborhoods in bringing the convergence of ideas, such as Kendall Square. Therefore, the influence of firms' and institutions' immediate environments (as represented by neighborhoods) on innovative activity has garnered growing attention in recent empirical research (Shearmur, 2012; Jang et al., 2017; Rammer et al., 2019; Roche, 2020; Matsiuk et al., 2024). Neighborhoods are also innovation environments, and the focus now shifts towards urban form, mobility, and third places, where users and knowledge workers interact. Coworking spaces, innovation hubs, cafés, squares, libraries, and universities are empirical proxies for "seed areas" where user communities, startups, civic tech groups, etc., meet and experiment, as the coworking and living-lab literature suggests (Capdevila, 2015; Roche et al., 2024; Howell, 2022; Kraus et al., 2022, Mariotti et al., 2021 & 2023). A study of coworking hubs finds that knowledge spillovers peak when start-ups are located within 20 meters of one another (Roche et al., 2024).

As with firms, the creation and sustainability of micro geographies of innovation depend on various factors shaped by local and global forces (Huggins, 2008). A competitive advantage based solely on one or two neighborhood determinants is possible, but it does not maximize innovative potential. Following Porter's Diamond (1990), which identifies four determinants of regional and national advantage, our study develops a conceptual framework that synthesizes local factors across four primary categories that may shape a neighborhood's innovation activity and, if improved, support its performance.

First, a) social and human resources reflect the availability of knowledge and skills - that is, the residents' ability to produce and absorb new ideas, while b) the economic features demonstrate the area's capacity to sustain and absorb productive and innovative activities, c) the morphology/environment acts as a setting for informal encounters and a variety of different uses; it creates the conditions for concentration, interaction, collaboration, knowledge diffusion, and the production of innovation d) land use, infrastructure shape the physical framework that enables accessibility and flows of people and activities, further supporting knowledge exchange and local innovation capacity, while e) mobility patterns determine how easily people can access places, services, and opportunities.

## 2.1. Research Aims, Questions & Empirical Strategy using Fine-Grained Data

When considering innovation capacity at the neighborhood or place level, the main assets to include in an empirical analysis are the non-locational and locational factors identified in the literature as enabling innovation activity. Our study identifies such assets to capture a range of dimensions of innovation performance and to offer a more holistic perspective on their overall impact on localities. It aims to unearth the factors that affect local innovation outcomes and provide policymakers, businesses, and researchers with insights into the evolving landscape of innovation. Most previous studies have examined relationships among some of the variables considered in this paper (Boschma, 2005; Cooke, 2001; Furman et al., 2002), but there is less empirical focus on assessing them collectively at the neighborhood level.

Our analysis centers on neighborhoods' innovation performance and also encompasses finer-grained factors related to socioeconomic status, as well as alternative features of neighborhoods' spatial structure. It acknowledges that each place has unique characteristics that should be supported to spur innovation; therefore, it aspires to provide actionable, bottom-up strategies to assess localities' strengths and harness their potential. It enables others to replicate the same process and identify significant innovation-enabling factors across diverse contexts. More specifically, it seeks to respond to the following questions:

**RQ1.** Are there individual or contextual-level characteristics that significantly affect a neighborhood's innovation performance?

**RQ2.** Are there key variables that are systematically significant between neighborhoods within the same state, but also between neighborhoods across different regions (e.g., Massachusetts vs. New York)?

To answer these questions, we apply machine learning algorithmic methods to identify key variables across the two US states, Massachusetts and New York. The empirical strategy focuses on traditional and alternative neighborhood assets, comprising 35 independent variables, as potential innovation-enabling factors. It examines the association between two indicators of innovation, a) the granted patents b) the start-up formation rate and socioeconomic attributes, but also newly introduced spatial variables capturing urban morphology, mobility patterns (commuting patterns to work), infrastructure, land use (e.g., the presence of local innovation centers, the presence of third spaces such as cafes and other formal and informal places of interaction) and urban morphological factors (e.g., the presence of parks and age of buildings).

Neighborhood characteristics were collected in 2012, and granted patents were collected with a four-year lag (2016), allowing for a robust assessment that mitigates simultaneity bias between innovation activity and contemporary spatial conditions. This approach provides sufficient time for innovation activities to materialize and be formally recorded.

## 3. Data & Measures

This section provides an outline of the selected data sources and datasets, as well as an overview of the variables incorporated to build an analytical database that reflects key aspects of local innovation systems and territorial characteristics. The study begins by defining the concept of a neighborhood with high innovation performance. It considers what it means (as an outcome) for a neighborhood to have high innovation performance, to identify significant independent variables that may influence the neighborhood's innovation outcomes across diverse neighborhoods.

Although our approach provides insights into potential associations, it does not establish causality, as the study is observational and does not manipulate variables to control for all possible confounding factors. The study leverages feature engineering techniques and machine learning models to examine how selected neighborhood assets contribute to variations in innovation performance. Traditional business and census data are combined with API-mined data (OpenStreetMap and Google Maps), the latter of which has not been widely used in the context of innovation geography. API-mined urban data are increasing rapidly, and their sources often provide access to finer-grained data at lower spatial scales, capturing diverse perspectives (Piovani et al., 2017).

All data are aggregated at the zip code level, providing a more localized view than city- or county-level data. Most countries use postal codes as a practical proxy for neighborhood analysis, as zip code data are widely available and commonly used by government agencies and data providers, making them easier to access and integrate across different sources. The construction of the database involves consolidating variables from multiple source datasets into a newly merged dataset that enables unified analysis and improved data accessibility. The merged database now provides a single, coherent data environment, reducing fragmentation and supporting more efficient querying, reporting, and analysis.

Each variable was first reviewed for consistency in format, definition, and measurement units before being harmonized and standardized, where necessary, for the analyses. This algorithmic methodology can measure performance and potential across ZCTAs, which were used as the initial unit, by merging census data with other datasets containing USPS zip code–based information. In most cases, the zip codes matched directly to ZCTAs; for our analysis, we retained all zip codes for which socioeconomic data were available. After the merge, the Massachusetts dataset included 495 zip codes, while the New York dataset included 1856 zip codes with census data, for a total of 2351 zip codes. The selected variables measured in total numbers were then normalized (per 1000 residents) to represent the relative distribution and intensity of the innovation hubs across urban/suburban areas, while for the predictive modeling, the independent variables selected were treated to ensure comparability and minimize numerical instability.

Overall, the key strength of this database and framework lies in its multidimensional measurement of a neighborhood's innovation performance, using independent data sources. For instance, we use OpenStreetMap to construct POI variables as indicators of urban vitality. However, these are not necessarily the most comprehensive, as previous studies have highlighted the limitations of voluntarily contributed data. Likewise, POI data capturing innovation spaces retrieved via Google Maps tend to be more complete in wealthier areas and urban cores, and less so in rural neighborhoods (Haklay et al., 2010).

While any single indicator has distinct limitations, combining multiple sources reduces the risk of bias, and multi-dimensional measurement represents an essential step toward more robust theory development and rigorous scientific analysis. The following figure outlines the methodology for constructing a database to support an exploratory analysis of potential determinants of innovation at the local level. The merged database now provides a single, coherent data environment, reducing fragmentation and supporting more efficient querying, reporting, and longitudinal analysis.

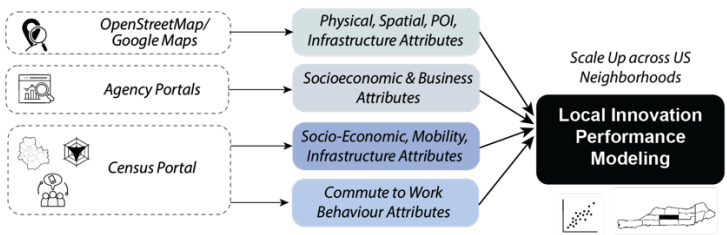

**Figure 1.** The proposed data integration process followed to create a local innovation performance model.

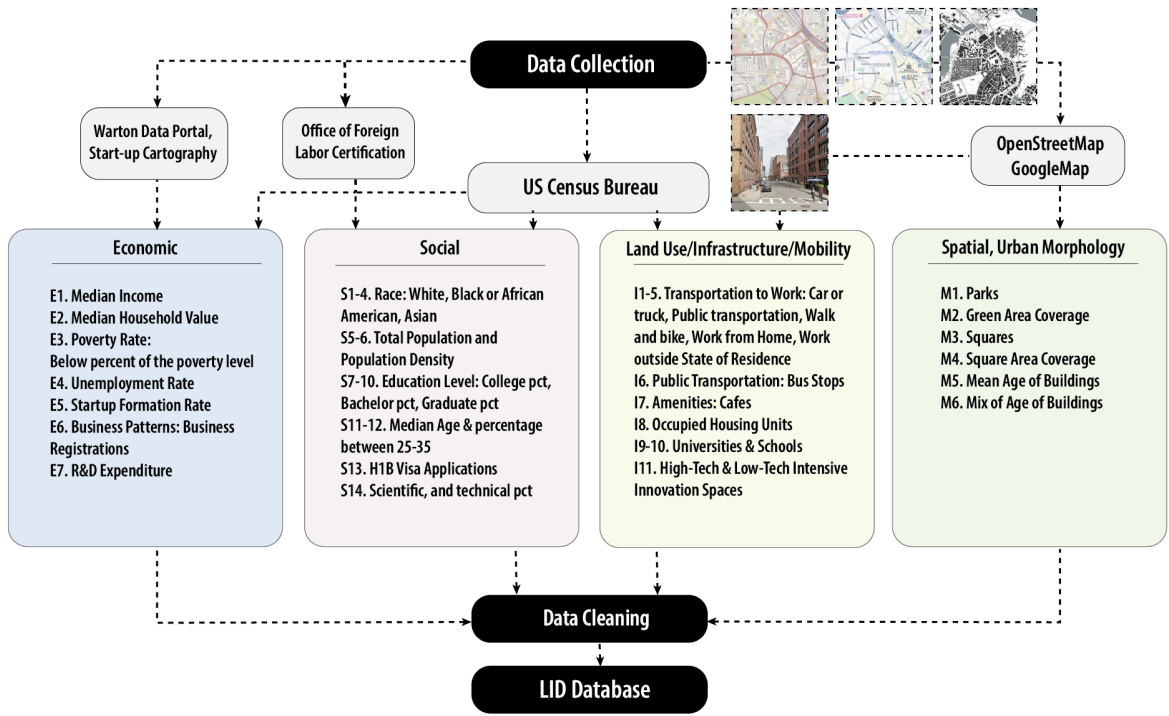

**Figure 2.** The methodology, data categories, sources, and the selected variables capturing neighborhood assets, including social, economic, mobility, spatial & urban morphological characteristics, authors' elaboration.

### 3.1. Patent records from USPTO as a traditional innovation indicator

The "Triple Helix" model emphasizes the synergies at the local level among universities, industry, and government in creating vibrant innovation ecosystems (Etzkowitz and Leydesdorff 1995). To capture knowledge exchange among universities, firms, and their ecosystems, researchers often rely on secondary data such as patent records, strategic alliance data, scientific co-publications, and information on R&D collaborations (Breschi and Lissoni 2009; Abbasiharofteh and Broekel 2020; Osland et al., 2022).

This study assesses how innovative a neighborhood is by capturing the number of granted patents. Data are retrieved from the USPTO, the most established source of information on innovation. Monitoring the number of patents granted by individuals and companies within the neighborhoods across New York and Massachusetts provides valuable insights into the neighborhood's capacity for technological advancement. The dataset is filtered to include all patents granted in New York and Massachusetts between January and December 2016. All entities have been geocoded using their exact geographic coordinates based on their filing location by zip code, and the geocoded locations of the granted patents have been filtered to the boundaries of the two states to exclude those outside the state boundaries.

Specifically, all records under the same RF ID were grouped, and for the zip code field, all unique zip codes were retained. This approach was particularly important for supporting spatial analysis, a critical focus of this study, as preserving spatial variability across distinct zip codes ensured accurate representation of geographic distributions and patterns (fig. A1-A4). The main limitation of this method is the potential for redundancy when multiple zip codes are retained for a single RF ID, which could complicate downstream analyses that require a single geographic reference per entity. For this study, we don't distinguish between corporate and individual records, as the primary objective is to identify the presence of granted patents in specific neighborhoods. By retaining at least one record per zip code for each patent, we ensure that every zip code involved in patenting activity is captured. This approach preserves critical geographic data and ensures that no zip Code associated with a granted patent is omitted from our analysis.

### 3.2 Start-up Cartography Project Dataset for Start-up Patterns

Innovation theories, such as the Knowledge Spillover Theory of Entrepreneurship (Acs et al., 2009), emphasize start-up activity as a critical mechanism for translating knowledge and ideas into marketable products and services, and the role of skilled labor in fostering innovation through knowledge creation and dissemination. Therefore, the start-up formation rate is pulled from the Start-up Cartography project (Guzman and Stern, 2020; Guzman et al., 2020), as a proxy that can provide crucial insights at the zip code level regarding the districts with higher start-up associated activity and growth (fig. A5-A8). While entrepreneurship is a key driver of economic development and social progress, only a small fraction of startups achieves the extraordinary levels of growth necessary to foster widespread economic dynamism (Guzman et al. 2020). This raises important questions for research, policy, and practice, particularly regarding the rate of startup formation, the underlying growth potential of startups, and the ultimate performance of startup cohorts across regions and over time.

The approach, first outlined by Guzman and Stern (2015) and later expanded upon by Guzman et al. (2020), introduced four key entrepreneurship statistics. By incorporating the Startup Formation Rate Index (SFR) into our database and employing it as an alternative measure of innovation output, we reflect entrepreneurial dynamics across Massachusetts and New York. This type of ecosystem metric facilitates comparison of entrepreneurial quality across the two U.S. states for the same period (2016) and better understanding of local innovation drivers and impacts. Following perspectives that view entrepreneurial activity as an outcome of innovative processes, our model conceptualizes the startup formation rate as an output of local innovation dynamics. Studies in the evolutionary and regional innovation literature argue that new firm formation reflects the commercialization of new knowledge and the emergence of market opportunities generated by prior innovative effort (Audretsch & Fritsch, 1994; Acs & Armington, 2006). In this view, entrepreneurial entry functions as a measurable indicator of the extent to which regions successfully convert knowledge inputs, such as R&D investments, skilled human capital, and technological capabilities, into economically relevant activity. Consequently, we treat startup formation not as a driver but as a dependent indicator of innovation performance, complementary to patent output in capturing the outcomes of regional knowledge creation for 2016 across both Massachusetts and New York.

### 3.3 Business Attributes from Agencies as established determinants of innovation performance

#### 3.3.1 Wharton Portal Dataset for Business Registration Records

Starting with the first category of independent variables integrated in the database, and more specifically, the business data, we collect data that reflects the concentration of businesses. Business registrations retrieved from the Warton Data Portal are used as an economic proxy to represent the clustering effect. The Wharton Business Registration Records dataset is a comprehensive database of registered businesses in the United States for the last decades. It provides detailed firm-level information, including company names, addresses, geographic identifiers, industry classifications, and financial estimates. The dataset is valuable for studying business formation, economic activity, firm growth, and the spatial distribution of enterprises. It includes both small and large businesses, tracking their establishment year, employment size, and estimated sales volume.

Researchers commonly use this dataset for analyses related to entrepreneurship, regional economic development, and market structure. In our study, we use this dataset to examine the concentration of business registrations at the zip code level in 2012 and in the regions under assessment. By leveraging its detailed firmographic and geographic data, we can examine the spatial distribution of businesses across the two U.S. states, identify geographic trends in business formation, and assess economic activity across neighborhoods. The incorporation of this large dataset from the Warton Data Portal enables a granular understanding of how business registrations evolve and vary by location. The concentration of business registrations is used as an economic proxy for clustering.

### 3.3.2 R&D expenditures – Using Wharton Portal Dataset

R&D expenditures in both public and private sectors are considered by innovation scoreboards to be an established innovation determinant, directly correlating with a region's ability to generate and absorb new knowledge. Several studies have already examined how private R&D expenditures contribute to this boost in innovation (Acs et al., 1994; Crescenzi et al., 2007; Furman et al., 2002). Of those, a substantial portion is empirical research supporting the hypothesis that R&D expenditures are essential to driving a firm's innovation activities (Stokey, 1995; Griliches, 1995; Bayoumi et al., 2000; Shefer and Frenkel, 1998; Frenkel et al., 2003).

Scholars have disaggregated R&D expenditures into three sub-categories: a) business enterprise; b) higher education; and c) government sector R&D (Rodríguez-Pose and Wilkie, 2019). From a more theoretical perspective, specific types of R&D expenditure are more closely associated with the generation of innovation than others (Malecki, 1991; Rodríguez-Pose, 1999). Business R&D, on the one hand, is more readily linked to the generation of "new goods and services, with higher quality of output and new production processes". In contrast, higher education and government sector R&D expenditure are more associated with advances in "scientific, basic knowledge and public missions" and the maintenance and expansion of the "stock of knowledge available for the society" (Guellec and van Pottelsberghe de la Potter, 2004). For our analysis, we incorporate firm-level annual R&D expenditure to capture patterns across all 2-digit NAICS sectors. Firm-level data are extracted from the WRDS (Wharton Research Data Services) portal, specifically from the Compustat North America database. This dataset includes financial and organizational information on publicly traded firms, with a particular focus on annual R&D (Research and Development) expenditures. It serves as the foundation for analyzing patterns and determinants of R&D investment across firms. Our study focuses exclusively on two variables: R&D expenditure (xrd) and firm location (addzip). Using this information, we compute the total annual R&D spending aggregated by zip code for 2012. This allows us to analyze the distribution of R&D investment based on firm-level spending patterns across diverse neighborhoods in the two states examined.

### 3.4 Attributes of individuals as innovation-enabling factors – independent or explanatory variables.

### 3.4.1 Capturing socioeconomic characteristics using census data

Diverse cultural backgrounds contribute to innovation by novel problem-solving approaches and the blending of ideas (Saxenian, 1994). These can be captured in the American Community Survey (ACS) 5-Year Estimates, conducted by the U.S. Census Bureau, which provides comprehensive and reliable demographic, social, economic, and housing data for small geographic areas, such as census tracts and zip codes. These estimates represent a rolling average of data collected over five years, offering a detailed and stable snapshot of local communities across the United States. The ACS 5-Year Estimates are particularly valuable for research and policymaking because they provide data at finer geographic levels than those available from the decennial census. They include variables such as population characteristics, income, education levels, employment status, commuting patterns, housing costs, unemployment rate, median income, and poverty rate, among others; they serve as proxies for residents' economic capacity. The level of granularity makes the 5-Year Estimates especially useful for analyzing smaller regions, such as neighborhoods or rural areas, where single-year estimates may not be statistically reliable due to small sample sizes. For example, the ACS 5-Year Estimates can be used to assess community needs, allocate resources, and plan infrastructure projects. They are widely utilized by researchers, urban planners, businesses, and government agencies to gain insights into local trends and disparities, supporting evidence-based decision-making.

For the zip-level estimates of the total population, White, Black, African American, and Asian populations from 2012, the estimate of the share of an educated population measured as the percentage of graduates, the rate of unemployment, median income, and poverty rate, American Community Survey (ACS) 5-Year Estimates are collected; these variables act as proxies of the residents' social characteristics and economic capacity. For population density, land-area data are obtained from the Tiger Line shapefile, along with the ACS 5-Year Estimates, to estimate the total population for 2012. Metrics such as the percentage of the population with tertiary education act as a proxy for a workforce capable of adopting and generating new technologies.

**3.4.2 H-1B employer data using Office of Foreign Labor Certification Dataset**

Given the fact that a substantial portion of the skilled workforce in the US tech/innovation clusters originates from overseas, for the analysis, the number of filed H1B applications is captured using data from the Office of Foreign Labor Certification (OFLC) website. H-1B Employer Data can be used to identify and analyze the geographic concentration of the foreign-born workforce across US innovation hubs. The Office of Foreign Labor Certification (OFLC) provides this publicly available dataset, offering a detailed view of employment-based visa petitions filed by U.S. employers. This proportion of a highly skilled foreign-born workforce serves as a proxy for the level of education and, simultaneously, cultural diversity in a neighborhood.

The H-1B program permits employers to hire foreign workers in specialty occupations that require advanced theoretical or technical expertise, such as those in IT, engineering, finance, healthcare, and academia. The dataset serves as a valuable resource for analyzing trends in the demand for skilled labor and understanding the role of global talent in the U.S. economy. The H-1B dataset includes certified, denied, and withdrawn Labor Condition Applications (LCAs) for 2012 for Massachusetts and New York, which are mandatory filings for employers seeking to hire H-1B workers. Key variables in the dataset include the employer's name, the job's geographic location (e.g., city, state, and zip code), and wage-related information, such as the offered salary and the prevailing wage for the position. Additionally, the dataset records the application's decision status—whether it was certified, denied, or withdrawn —and the North American Industry Classification System (NAICS) code, which identifies the employer's industry sector.

**3.5 Identifying infrastructure and land use as innovation-enabling factors**

**3.5.1 Identifying innovation spaces through Google Maps API**

Incubators and accelerators provide high-tech support and resources for early-stage companies, while designated innovation centers and hubs complement the local innovation ecosystems by supporting low-tech, more informal, and social innovation-related activities. Both types of spaces have become significant contributors to the local innovation ecosystems. As part of a new exploration of the innovation capacity of localities, keywords such as accelerator, co-working space, incubator, etc., are used for searching spaces, which can serve as enablers for innovation activity across New York's and Massachusetts' zip codes, so that a comprehensive understanding of the dynamic landscape of innovation spaces across these two states is gained.

To do this, Application Programming Interfaces (APIs) are used as a robust tool to pull the locations of innovation spaces. More specifically, all entities have been geocoded to their exact geographic coordinates based on their names using the Google Maps Geocoding API, and the geocoded locations of the innovation spaces have been filtered by zip code in the two states. As part of a new exploration of the innovation capacity, keywords for place name such as a) accelerator, b) co-working space, c) incubator, d) innovation center, e) innovation hub, f) innovation park, g) start-up, h) tech hub, i) technology park are used for searching spaces, which provide support and resources for early-stage companies, and serve as enablers for innovation activities across New York's and Massachusetts' zip codes. Innovation spaces serve as an additional proxy for capturing innovation activities across the two US states and for identifying trends in both urban and suburban contexts, within traditional and emerging innovation hubs.

**3.5.2 High and low-tech knowledge-intensive spaces as innovation enabling factors - using OSM API**

The study by Dubey et al. (2020) highlights the potential of combining OpenStreetMap (OSM) data with machine learning (ML) techniques to generate high-resolution socio-economic indicators, which can serve as proxies for understanding economic development. The authors demonstrate that ML models trained on OSM-derived features, such as road density, building footprints, and the distribution of amenities, can accurately estimate key economic metrics, including income levels, urban development patterns, and infrastructure accessibility. Our study extends this scope and aims to use these metrics to provide valuable insights into whether areas with well-developed infrastructure, diverse land use, efficient transportation networks, and higher concentration of open space exhibit stronger innovation dynamics. In the end, our findings can underscore the effectiveness of open geospatial data in supporting policymakers and researchers in identifying early-stage innovation hotspots, guiding urban planning, and fostering regional economic growth.

Therefore, the locations of universities and schools for formal interactions are retrieved from the OpenStreetMap API, as these spaces provide a stable framework for innovation through governance, regulation, and support systems (Rodrik et al., 2004). University and school tags are used to collect their locations within the two states' zip codes to support analysis of the enabling factors. A key feature of a technology cluster is its micro-geography, as being in a specific micro-location within the cluster allows firms to engage in and gain from localized, highly specialized knowledge exchanges that happen exclusively through face-to-face interactions and unexpected encounters (Petruzzelli et al., 2007; Broström, 2010; Balland, 2012; Cassi and Plunket, 2014; Steinmo and Rasmussen, 2016).

Building density and diverse land use enhance these face-to-face interactions and create an urban buzz, further increasing innovation productivity (Storper and Venables, 2004; Wood and Dovey, 2015). Therefore, café locations are retrieved via the OpenStreetMap API and serve as a proxy for the presence of 'third spaces', which strongly enhance knowledge-based economic vitality and entrepreneurship (Choi et al., 2024). Using OpenStreetMap (OSM) as a source for identifying the 'third' spaces of encounter is an effective way to address the challenge of conducting assessments at the zip code level across entire U.S. states. OSM helps get a more granular view of local public and semi-public spaces and the concentration of amenities.

Additionally, a variable related to public transportation, such as the number of bus stops for each zip code is collected to understand the ease of interaction and access to and from the neighborhoods. This variable is used to understand the effects of public transportation on innovation performance and address challenges related to urban sprawl, congestion, and inequality. Incorporating variables related to transportation infrastructure is crucial for understanding the quality, functionality, and sustainability of urban environments. Previous studies have examined the significance of infrastructure in enabling direct innovative endeavors (Kerr et al., 2014; Fehder et al., 2014).

### 3.6 Alternative enabling factors: The in-between space - mobility, spatial and urban morphology

#### 3.6.1 Commuting patterns as a proxy of innovation networks using census data

Mobility is fundamental to the communication, networking, and interaction among innovation agents, as suggested by theories of innovation ecosystems and spatial dynamics (Cooke et al., 1997; Bathelt et al., 2004); it facilitates knowledge transfer and collaboration, crucial for innovation. It is a key enabler of both communication and network formation; evidence suggests that innovative firms in the U.S. are more likely to establish themselves in neighborhoods characterized by walkability and high public transit use (Hamidi and Zandiatashbar, 2018). These measures reflect the ease of access and interaction.

Empirical studies indicate that greater land-use diversity and denser street intersections are positively associated with walking and cycling activity. Block size, as assessed using GIS polygon databases by Long and Huang (2019), is negatively related to urban economic vitality. Comparable findings appear in transport research, where smaller blocks have been found to encourage walking and cycling (Ewing and Cervero, 2010). Our study uses mobility behavior data at the zip code level to capture commuting patterns to work, based on the American Community Survey (ACS) 5-Year Estimates for 2012, retrieved from the U.S. Census Bureau, for the zip codes of two U.S. states (Massachusetts and New York). Despite the limitation of the ACS) 5-Year Estimate that previously discussed, they include variables that capture critical information on commuting patterns to work either a) walking, b) biking, c) using the car or van, d) using public transportation, or e) working from home, f) working from another state g) working in state of residence h) working in county of residence i) working outside county of residence j) working outside state of residence k) working in place of residence.

For our analysis, variables are transformed into percentages based on the total population per zip code, and the percentages of workers aged 16 and older who commute by walking or biking are combined into a single variable. In combination with the variable introduced earlier, which measures the locations of bus stops captured via OpenStreetMap APIs, these proxies are used to examine the association between public transportation, walkability, ease of access, and innovation performance. Innovative firms in the U.S. commonly establish themselves in walkable neighborhoods with higher transit use (Hamidi and Zandiatashbar, 2018). Therefore, assessing both commute behavior patterns and infrastructure is crucial for understanding the quality, functionality, and sustainability of urban environments.

### 3.6.2 Urban morphological characteristics using OpenStreetMap

Analyzing morphological assets, such as the number of public squares and parks, can be used as another proxy to capture informal interactions. These elements were collected through OpenStreetMap APIs, as other proxies for places that facilitate informal interactions. Previous studies have highlighted that the physical layout is critical to understanding neighborhood capacity (Roche, 2020). In our study, the number of squares and parks is used to assess the distribution of green and open spaces, the livability of neighborhoods, the walkability to open spaces, and access to informal interactions. Together, these metrics help to understand the urban morphology of the assessed zip codes in the two U.S. states. These variables, retrieved from OSM, also provide valuable insights into urban and environmental planning. Park and square land area (measured in acres) captures the proportion of land dedicated to greenery, relative to the population of the zip code (normalized per 1000 residents), indicating the accessibility of nature for the resident population.

Measuring environmental and urban morphological assets involves evaluating the interplay among components, such as streets, buildings, and open spaces, to understand how effectively they support innovation activities and contribute to overall neighborhood performance. This assessment considers factors such as accessibility, environmental impact, and aesthetic appeal, which can enable collaboration, enhance the proximity of innovators, and create a critical mass. While some of these spatial indicators have been examined in previous studies, they have typically been analyzed in isolation, focusing on the individual effects of specific variables. At the same time, we identify a gap in empirical studies: the need to systematically integrate additional spatial indicators into innovation geography research. Therefore, our study applies a holistic approach, necessary to capture the complex, interrelated dynamics, including urban morphological aspects that collectively enable local innovation performance.

### 3.6.3 Building age and mix as potential innovation enabling factor - using census data

Jacobs (1961) contended that large-scale, centrally planned redevelopment projects fail because they overlook the reality that vibrant cities are not engineered but evolve organically, primarily through messy, spontaneous, and complex processes. From this perspective, neighborhoods that have developed gradually rather than through rapid, intensive redevelopment are likely to show a healthier mix of land uses, with more walkable environments, fewer vacant spaces, and greater functional diversity. Older buildings, empirically associated with higher levels of social interaction and greater housing affordability (Montgomery, 2013), can be efficiently assessed at scale through municipal property databases.

Building age has been quantified in previous studies, using two indicators: the mean building age and the standard deviation of building age (Huang et al., 2023). This follows Jacobs' principle that "the district must mingle buildings that vary in age and condition, including a good proportion of old ones" (1961, p. 187), as supported by data on building ages obtained from government sources (HKHS, 2016). In the context of our local innovation study, the age of development can indicate building conditions, help identify strengths and weaknesses within the urban fabric, and reveal the historical and physical characteristics of urban infrastructure. Jacobs (1961) argued that including older structures is essential for fostering innovation as "cities need old buildings so badly" and "new ideas must use old buildings", which provide the experimental space needed.

Our study analyzes these aspects by measuring two variables, estimated from the 2012 census data, using the ACS 5-Year Estimates at the zip code level. Despite their long-standing importance and popularity, Jacobs's theories have not been tested at this level of granularity to examine the relationship between urban fabric and innovation activities. Our analysis helps assess whether there is an association between older developments and designs, the diversity of building ages in the examined geographic area, and higher concentrations of innovation activities. These two variables serve as proxies for understanding the relationship between urban form and the diversity of the urban fabric and innovation activities, as they were previously identified as encouraging a broader range of economic and social activities (Jacobs, 1961).

### 4. Understanding dependencies between neighborhoods' assets and innovation performance

Our study first employs a random forest algorithm as part of the analytical process to identify the most important of the 35 variables for predicting the innovation output in our model. This algorithm, known for its effectiveness in handling complex datasets and reducing overfitting, helps determine which factors most influence the dependent variables. By leveraging this approach, model-building efforts are streamlined and focused on the key drivers that better predict the concentration of granted patents and the startup formation rate. Among the selected variables, fewer social, economic, infrastructure, and urban morphological characteristics are expected to have stronger associations with innovation outcomes, as measured by granted patent records (Table A3).

More specifically, the feature-importance attribute estimates each feature's importance by measuring how often it is used to split the data and how much it reduces impurity in the resulting subsets. A Random Forest Regressor with 1000 trees was trained on the dataset eight times, each with a different random seed. Feature importance was computed for each run using the mean decrease in impurity, and the results were then averaged across runs. Features were then ranked by their average importance to identify the most influential predictors. When examining the number of patents granted separately for New York State and Massachusetts, we find that different neighborhood characteristics have significant predictive value for the dependent variable.

For example, when running the model for **Massachusetts**, we see that the H1B visa applications, the innovation spaces, the R&D expenditure, and the cafes per 1000 residents as well as the median income, the percentage of residents with bachelor degree, along with business registrations per 1000 residents, parks and squares per 1000 residents have the most significant effect on the predicted innovation outcome in its neighborhood's (fig. 12), while when we run the model for all the zip codes of **New York State,** we find that R&D expenditure per 1000 residents, the number of H1B visa applications per 1000 residents, the bus stops per 1000 residents, and the number of business registrations, percentage of residents between 25 to 34 years old, the cafes per 1000 residents along, the business registrations along with the with population with graduate degree, are the most significant features for the model (fig. 13).

Overall, when analyzing all zip codes from both states collectively, we observe that the number of H1B applications per 1000 residents, the R&D expenditure per 1000 residents, the percentage of people in the age of 25 to 24 years, the business registrations per 1000 residents, the number of bus stops and cafes per 1000 residents, the population density, the percentage of people obtaining a graduate degree have the most significant impact on the dependent variable, the number of granted patents per 1000 residents (fig.14).

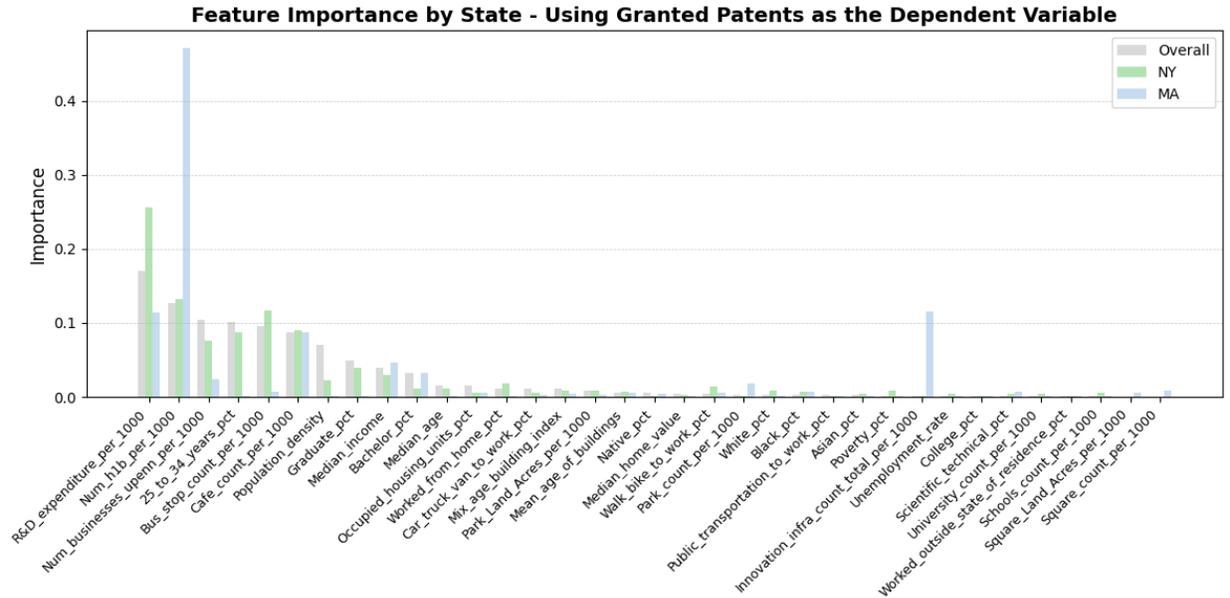

**Figure 11.** Random forest algorithm for assessing feature importance in innovation outcomes, using granted patents as the output for all zip codes for Massachusetts and New York State respectively, authors' elaboration.

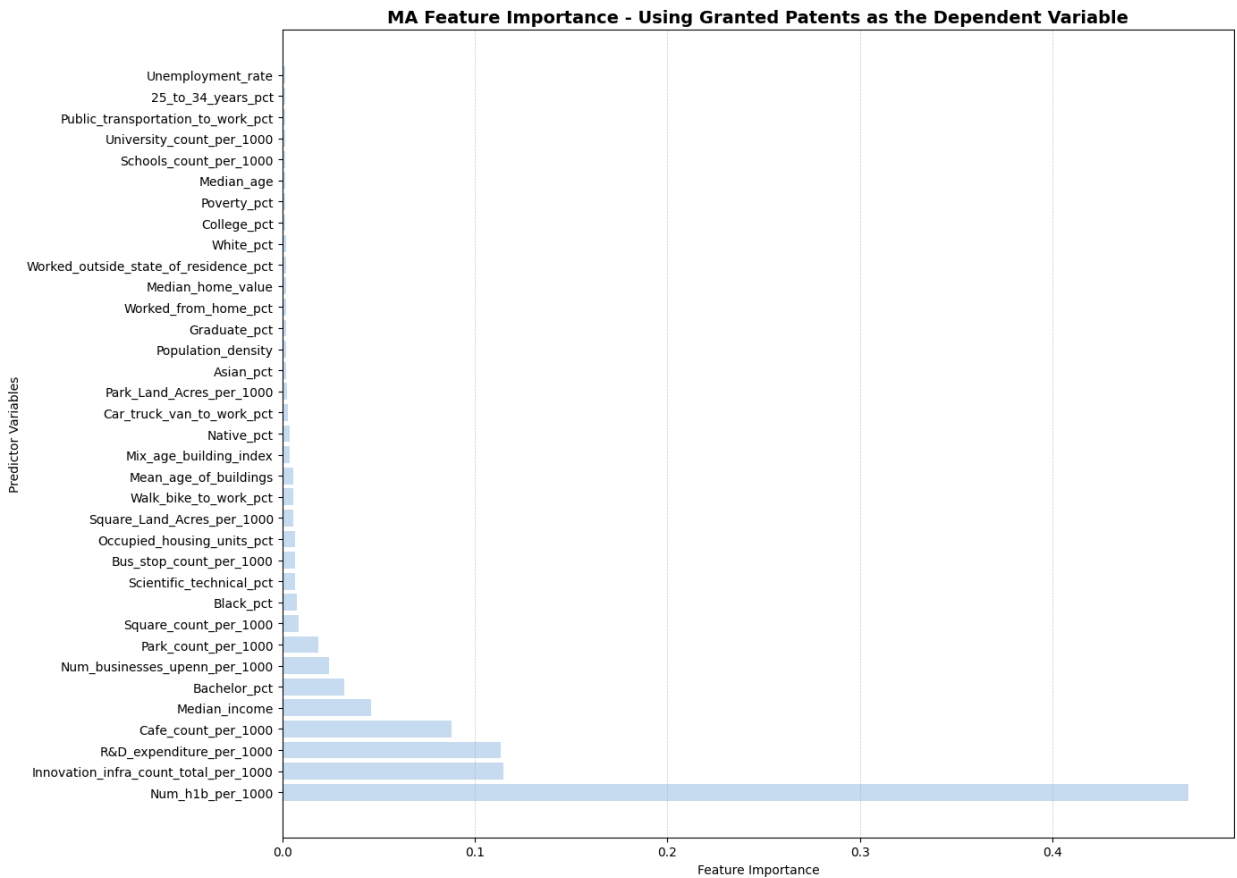

**Figure 12.** Random forest algorithm for assessing feature importance in innovation outcomes, using granted patents as the output for all zip codes for Massachusetts, authors' elaboration.

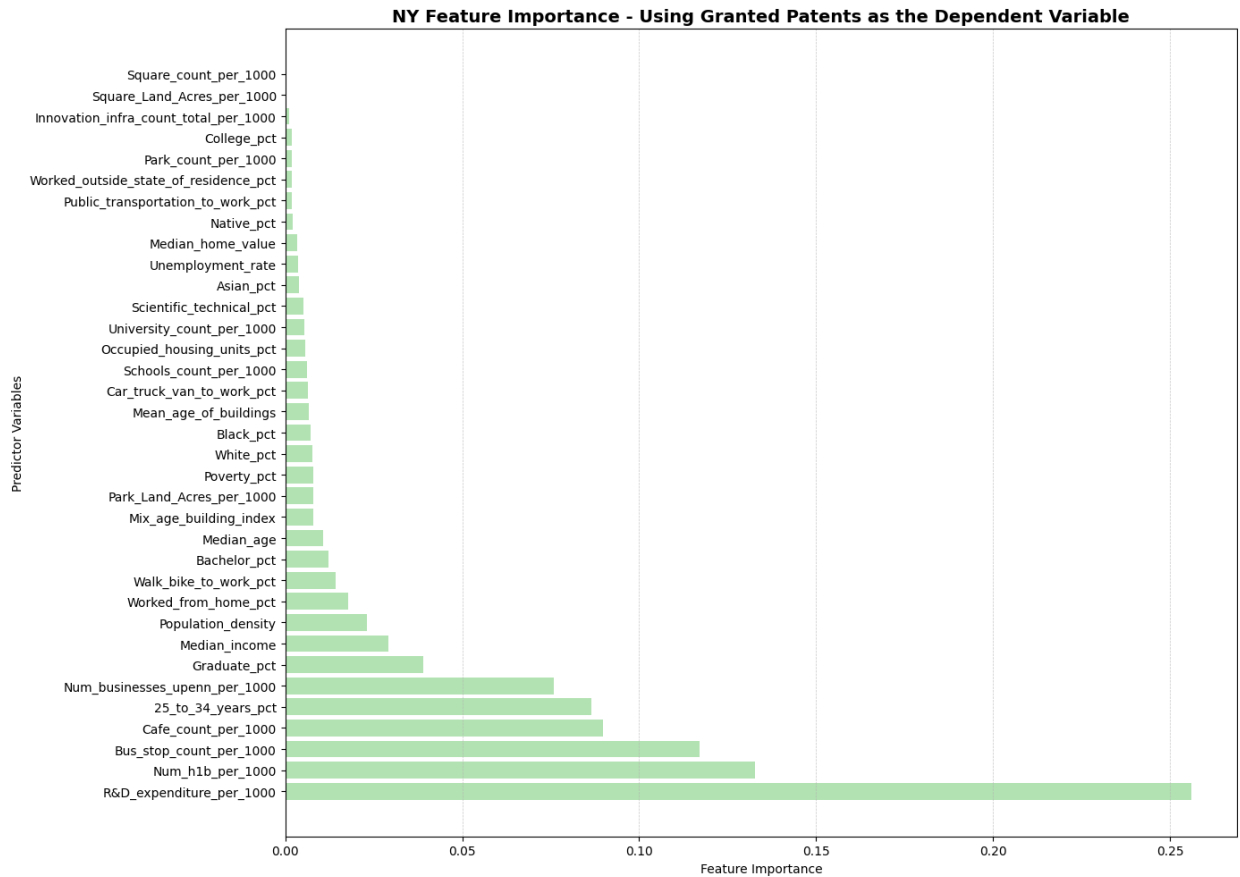

**Figure 13.** Random forest algorithm for assessing feature importance in innovation outcomes, using granted patents as the output for all zip codes for New York State, authors' elaboration.

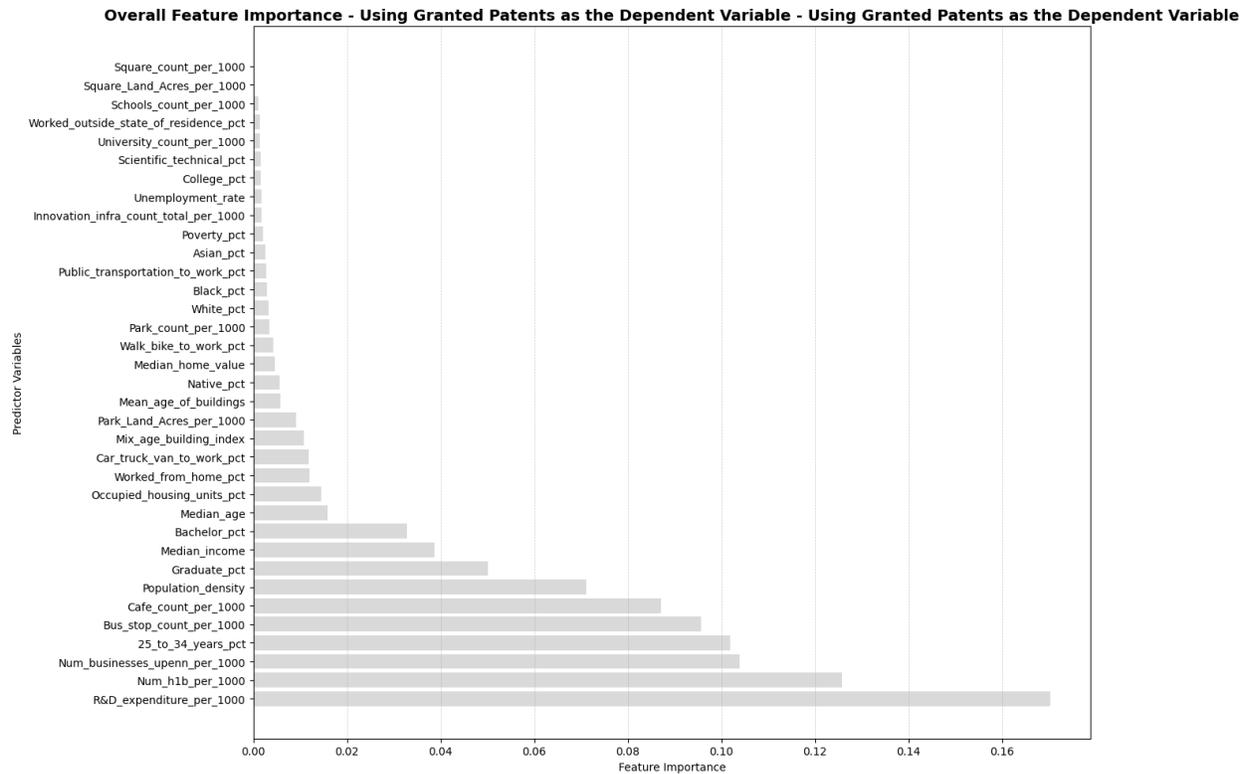

**Figure 14.** Random forest algorithm for assessing feature importance in innovation outcomes, using granted patents as the output for all zip codes for Massachusetts and New York State together, authors' elaboration.

For our analysis, we repeated the same process, with the start-up formation rate as the outcome (Table A4). When running the model for **Massachusetts**, we find some of the significant variables for granted patents model (H1B visa applications the number of R&D expenditure, and number of business registrations, cafes per 1000 residents and percentage of residents in age group 25-34, with the addition of the percentage of occupied housing units in the zip code, the median home value, percentage of professionals in scientific technical fields and squares per (fig. 16), while when we run the model for all the zip codes of **New York State,** the percentage of occupied units, and median home value, the Percentage of Asian population, percentage of people with college degree, the number of H1B applications per 1000 residents, the percentage of people using transportation to work, business registrations and population density, the percentage of people walking/biking to work appear to be the most significant for the model (fig. 17).

Overall, when analyzing all zip codes from both states collectively, we observe that percentage of occupied units is the most significant feature, followed by median home value, population density, the percentage of people using public transportation, number of H1B visa applications per 1000 residents, the business registrations per 1000 residents, the percentage of Asian population, and residents worked outside the state of residence, the percentage between 25-34 years old, have the most significant impact on the dependent variable, the number of start-up formation rate per 1000 residents (fig. 18).

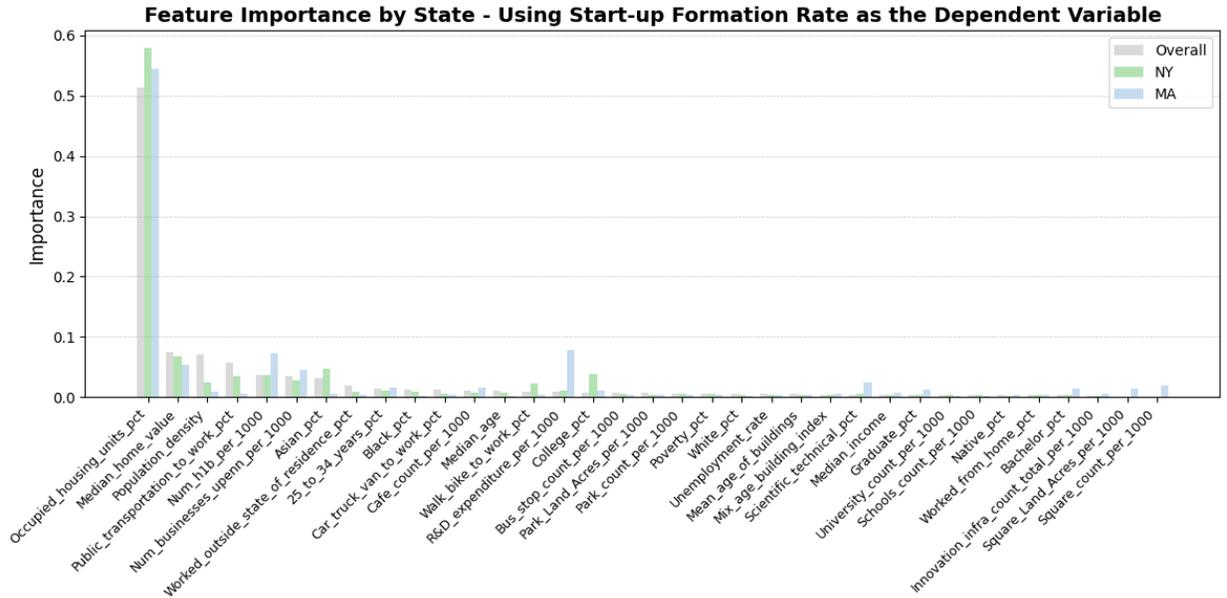

**Figure 15.** Random forest algorithm for assessing feature importance in innovation outcomes, using start-up formation rate as the output for all zip codes for Massachusetts and New York State respectively, authors' elaboration.

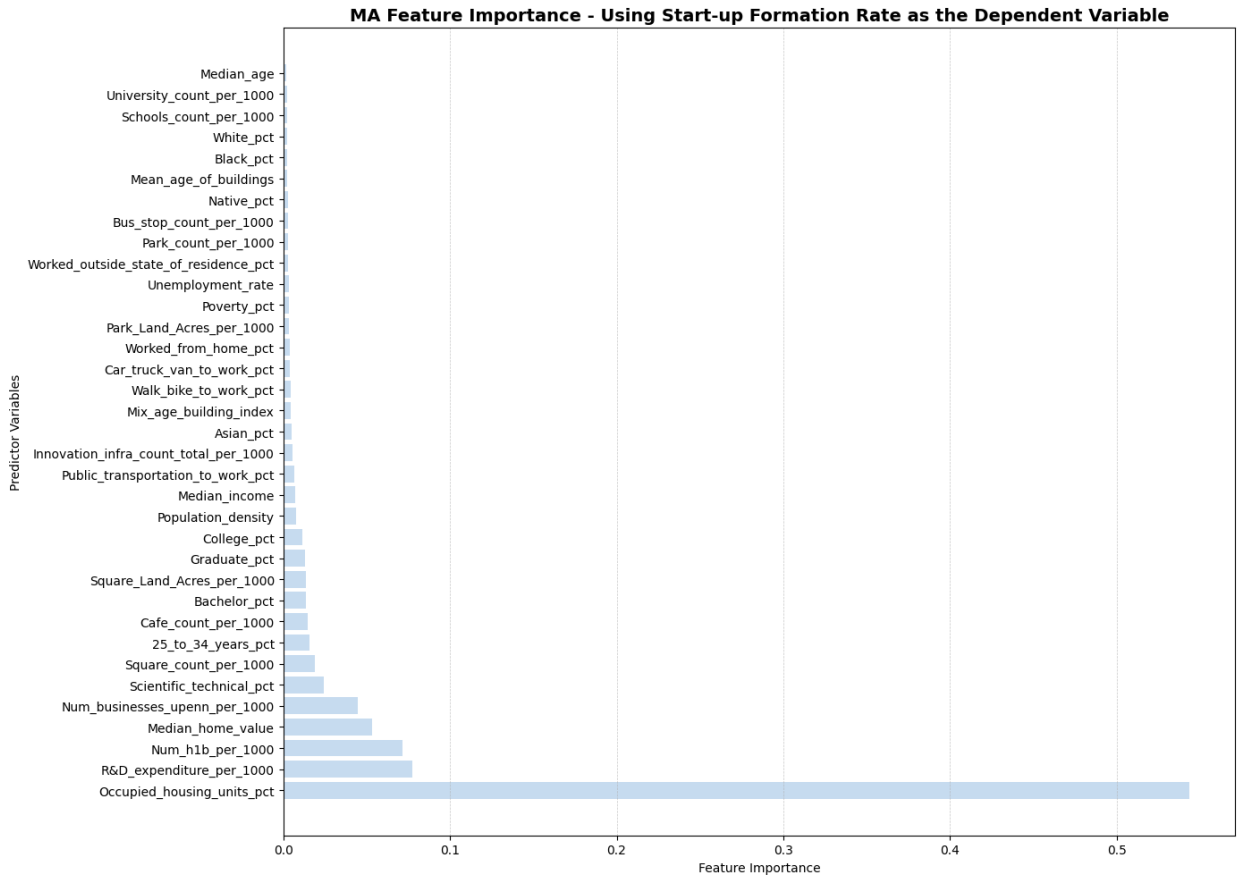

**Figure 16.** Random forest algorithm for assessing feature importance in innovation outcomes, using start-up formation rate as the output for all zip codes in Massachusetts, authors' elaboration.

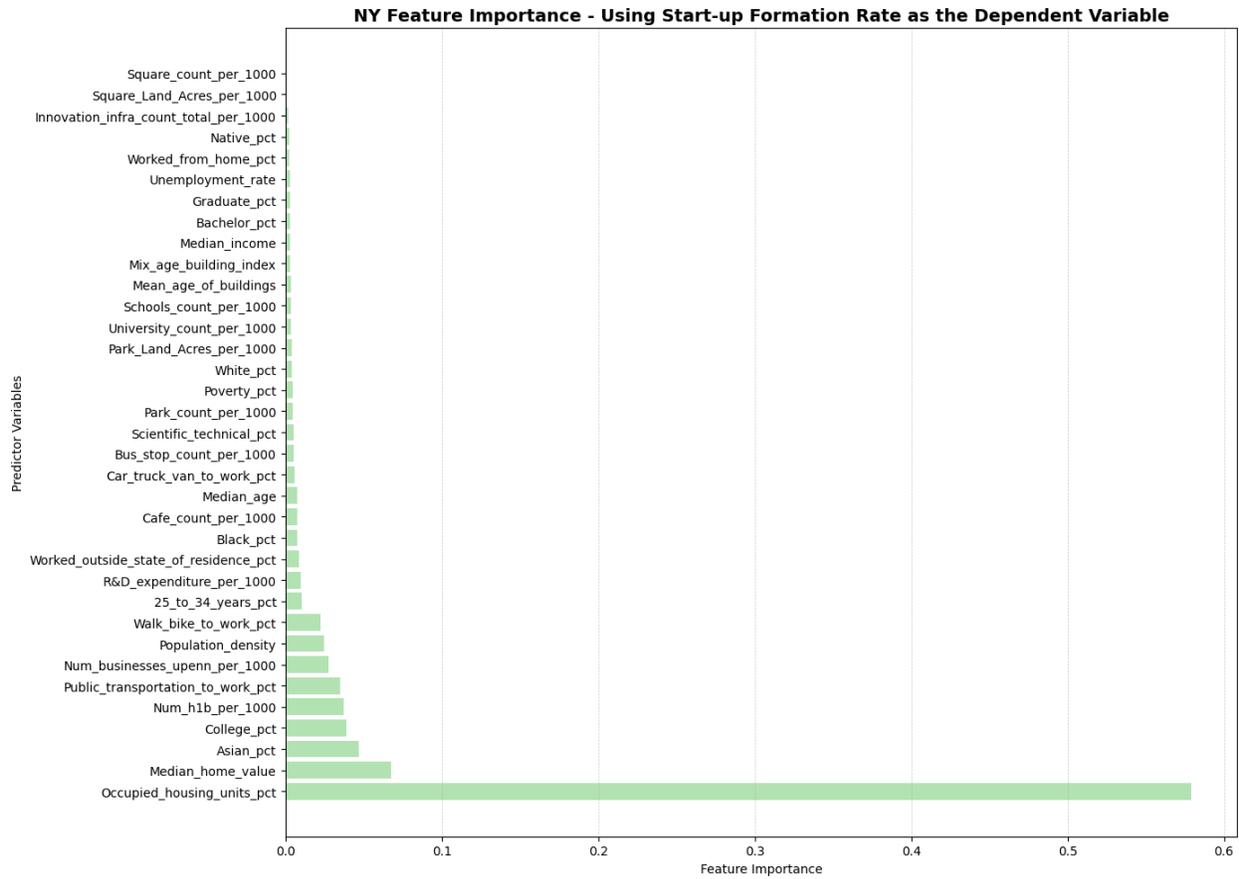

**Figure 17.** Random forest algorithm for assessing feature importance in innovation outcomes, using start-up formation rate as the output for all zip codes for New York State, authors' elaboration.

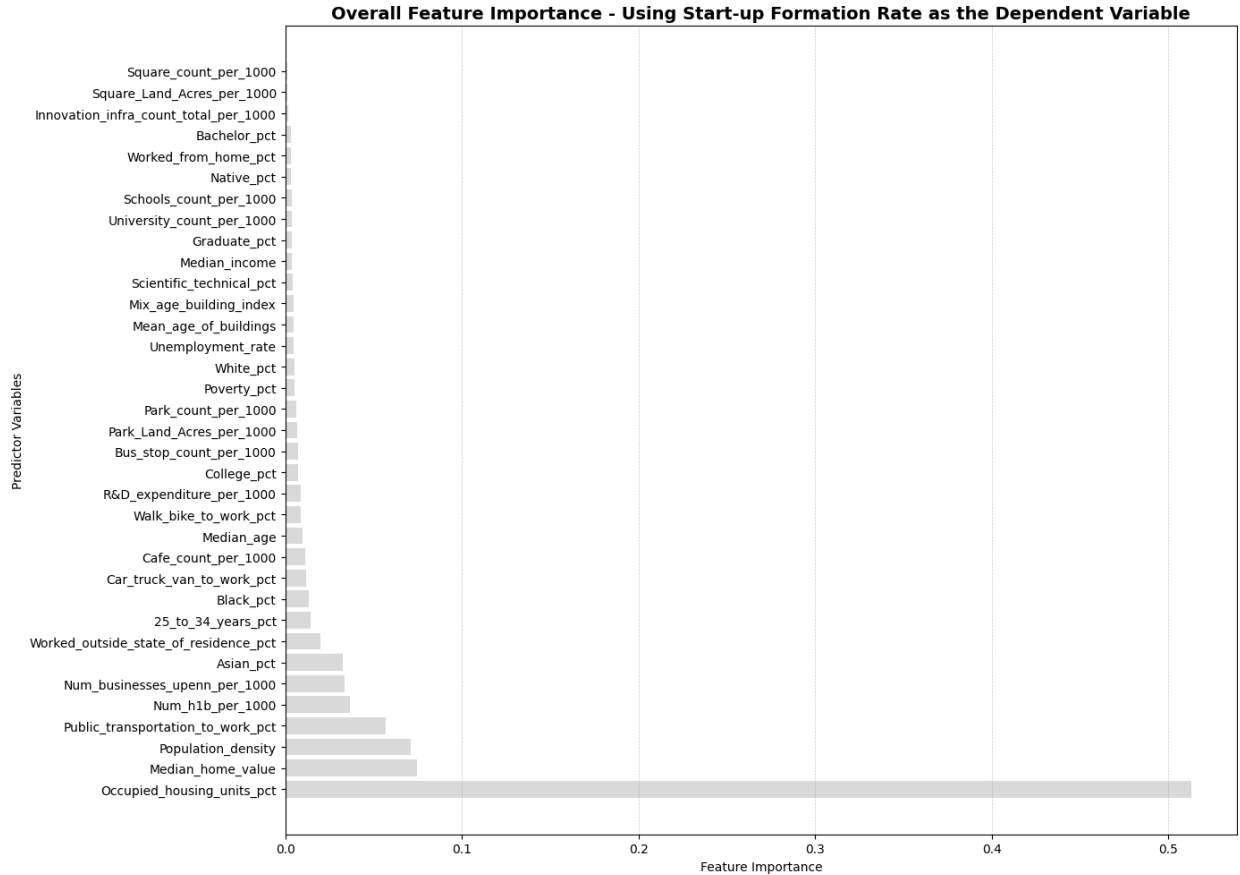

**Figure 18.** Random forest algorithm for assessing feature importance in innovation outcomes, using start-up formation rate as the output for all zip codes for Massachusetts and New York State together, authors' elaboration.

## 5. Conclusions, Limitations, and Future Outlook

In this paper, a new framework is introduced to identify potential innovation-enabling factors at a lower-than-city spatial scale across diverse US regions. For this, a new database is constructed by merging data across different aspects of a neighborhood: social characteristics, economic and land-use context, urban mobility, morphology, and infrastructure to understand innovation outcomes. First, the RF algorithm's capacity and correlation matrices are deployed to assess the influence of each neighborhood feature on the innovation performance measured in a) granted patents, b) startup formation rate. It shows that a few features, such as H-1 B visa applications, business registrations, and cafes per 1000 residents, have stronger associations with innovation outcomes across both states, and when tested with both outputs.

However, the order of features differs when innovation outcomes are predicted using the two dependent variables. Features related to housing occupancy and value are substantially more significant when the start-up formation rate is used as the predicted variable. Overall, both models provide evidence that features describing populations with origins overseas (H1B visa applications) and the Asian population are significant in accordance with previous studies (Saxenian, 1994) that discussed how immigrant diversity, especially highly skilled engineers from China, India, and other countries, significantly strengthened Silicon Valley's innovative capacity.

Furthermore, our database incorporates a comprehensive set of control variables known to influence innovation (socioeconomic characteristics). It introduces new variables reflecting land use and urban morphology to explore their association with innovation activities at the zip level. Our study acknowledges that, because several of the in-cluded variables, such as H-1B visa applications and business registrations, are conceptually and empirically correlated, there is a risk of confounding and inflated variance estimates. To address this, diagnostic tests for multicollinearity should be conducted as the next step in this study, including Principal Component Analysis (PCA) and additional correlation analyses. Additionally, given the fine-grained spatial scale of the data, our study acknowledges the likelihood of spatial autocorrelation, whereby nearby observations are not entirely independent. This potential clustering effect will be accounted for in a future study. Employing techniques, Moran's I analysis would help address this.

After this study, we better understand the relationships between socioeconomic, urban morphological, and infrastructure characteristics and their innovative outcomes. The estimates provide significantly greater granularity than prior estimates in the literature and suggest that addressing the levels of the significant variables examined in this study is essential for understanding local-level innovation. The results provide strong evidence that specific socioeconomic, infrastructure, and spatial variables may have a more substantial effect on innovation performance. This exploratory geospatial data analysis emphasized the role of big urban data analytics in enabling fine-grained decision-making at finer spatial scales within both urban and suburban ecosystems. By tracking the real-time growth of specific variables, without waiting for innovation activity to occur, policymakers and researchers can monitor neighborhood outcomes to inform local innovation policy.

Since there is a strong indication that specific spatial characteristics, such as the park land in a neighborhood, are enabling factors of innovation outputs (quantified both with granted patents and startup formation rate), our next goal is to develop predictive modelling that can fully capture the effects of these independent variables in more depth. Going forward, the study also aims to expand geographically and include additional indicators to examine whether other neighborhood characteristics drive innovation. Additionally, we strive to provide an open data infrastructure that enables others to build on our measures and to identify how the project's findings can inform policymaking and current research in the field. It would also be important, as the next step of this study, to determine whether the same variables affect innovative outcomes across U.S. states with diverse socioeconomic contexts (West vs. East Coast). By applying the model to several neighborhoods in the US, its validity in environments with district-level neighborhood characteristics can be assessed.

# Appendix

**Massachusetts State Statistics**

**Table A1**. Summary Statistics of Neighborhood Elements for the State of Massachusetts.

| Variable | Median | Mean | SD | Group |
| --- | --- | --- | --- | --- |
| H1B Applications per 1000 residents | 0.21 | 1.552 | 9.203 | Social |
| Scientific Technical pct | 0.12 | 0.131 | 0.074 | Social |
| White Pct | 8655.0 | 12493.96 | 12191.66 | Social |
| Black Pct | 0.929 | 0.871 | 0.15 | Social |
| Native Pct | 0.014 | 0.04 | 0.085 | Social |
| Asian Pct | 0.0 | 0.002 | 0.005 | Social |
| 25 to 34 Years Pct | 0.018 | 0.04 | 0.059 | Social |
| College Pct | 0.001 | 0.006 | 0.032 | Social |
| Bachelor Pct | 0.216 | 0.207 | 0.078 | Social |
| Graduate Pct | 0.225 | 0.228 | 0.089 | Social |
| Population Density | 0.149 | 0.18 | 0.123 | Social |
| Median Age | 293.961 | 1185.146 | 2433.456 | Social |
| Median Income | 42.2 | 42.054 | 8.037 | Economic |
| Unemployment Rate | 33892.0 | 34929.562 | 11612.469 | Economic |

| Variable | Median | Mean | SD | Group |
|---|---|---|---|---|
| Poverty Pct | 0.074 | 0.08 | 0.04 | Economic |
| Median Household Value | 323600.0 | 353692.872 | 146844.545 | Economic |
| Business Registrations per 1000 residents | 0.084 | 1.495 | 8.119 | Economic |
| R&D Expenditure (Million) per 1000 residents | 3390.0 | 4814.125 | 4749.455 | Economic |
| Occupied Housing Units Pct | 0.491 | 0.709 | 0.817 | Infrastructure |
| Schools per 1000 residents | 0.981 | 1.418 | 1.634 | Infrastructure |
| Universities per 1000 residents | 42.816 | 63.448 | 127.975 | Infrastructure |
| Mean Age of Buildings | 48.737 | 57.161 | 124.989 | Infrastructure |
| Mix Age of Buildings | 0.847 | 0.793 | 0.146 | Infrastructure |
| Innovation Spaces per 1000 residents | 0.069 | 0.194 | 0.465 | Infrastructure |
| Parks per 1000 residents | 0.659 | 1.045 | 1.416 | Urban Morphology |
| Squares per 1000 residents | 0.11 | 0.311 | 0.483 | Urban Morphology |
| Park Land Acres per 1000 residents | 4.217 | 30.659 | 365.809 | Urban Morphology |
| Square Land Acres per 1000 residents | 0.01 | 0.189 | 0.587 | Urban Morphology |
| Car Truck Van to Work Pct | 0.883 | 0.827 | 0.167 | Urban Mobility |
| Public Transportation to Work Pct | 0.021 | 0.056 | 0.088 | Urban Mobility |
| Walk Bike to Work Pct | 0.023 | 0.053 | 0.098 | Urban Mobility |
| Worked from Home Pct | 0.046 | 0.055 | 0.053 | Urban Mobility |
| Worked Outside State of Residence Pct | 0.025 | 0.043 | 0.054 | Urban Mobility |
| Bus stops per 1000 residents | 1.337 | 2.387 | 3.09 | Urban Mobility |
| Patents per 1000 residents (2016) | 0.224 | 2.431 | 10.998 | Innovation Outcomes |
| SFR (2016) | 33.0 | 52.393 | 56.538 | Innovation Outcomes |

*Note: This table reports summary statistics for neighborhoods' elements zip-year socio-economic observations retrieved from the Census Bureau in 2012 for zip codes across Massachusetts. There are 495 such pairs in the dataset; for this analysis, an inner join is used to retain only zip codes retrieved from the census dataset. Population density is calculated by dividing the population by the land area of the respective zip code, then multiplying by 1000000 to convert to **people per square mile.** The variables under the "Infrastructure" group and the rows under "Urban Mobility & Spatial Characteristic " present morphological characteristics, such as parks and squares as well as mobility patterns to work, which have been retrieved more recently from OpenStreetMap, similar to the innovation spaces retrieved from the Google map; however, they have been used as the best approximation for the 2012.*

**New York State Statistics**

**Table A2.** Summary Statistics of Neighborhood Elements for New York State.

| Variable | Median | Mean | SD | Group |
|---|---|---|---|---|
| H1B applications per 1000 residents | 0.0 | 1.649 | 48.665 | Social |
| Scientific technical pct | 0.084 | 0.09 | 0.067 | Social |
| Total population | 3705.5 | 11556.844 | 17516.51 | Social |
| White pct | 0.937 | 0.848 | 0.209 | Social |
| Black pct | 0.012 | 0.062 | 0.137 | Social |
| Native pct | 0.0 | 0.005 | 0.037 | Social |
| Asian pct | 0.009 | 0.036 | 0.072 | Social |
| 25 to 34 years pct | 0.002 | 0.023 | 0.517 | Social |
| College pct | 0.232 | 0.228 | 0.089 | Social |
| Bachelor pct | 0.146 | 0.165 | 0.104 | Social |
| Graduate pct | 0.099 | 0.128 | 0.101 | Social |
| Population density | 100.354 | 2149.133 | 6040.969 | Social |
| Median age | 41.6 | 41.747 | 8.014 | Economic |
| Median income | 28390.0 | 31180.628 | 12361.922 | Economic |
| Unemployment rate | 0.076 | 0.085 | 0.067 | Economic |
| Poverty pct | 0.094 | 0.118 | 0.105 | Economic |
| Median home value | 171500.0 | 255169.816 | 194951.905 | Economic |
| R&D expenditure per 1000 residents | 0.078 | 2.525 | 26.427 | Economic |
| Occupied housing units pct | 1444.0 | 4295.945 | 6452.737 | Infrastructure |
| Schools per 1000 residents | 0.403 | 0.654 | 1.567 | Infrastructure |
| Universities per 1000 residents | 0.805 | 1.308 | 3.135 | Infrastructure |
| Business registrations per 1000 residents | 36.982 | 106.277 | 1877.635 | Infrastructure |
| Mean age of buildings | 49.525 | 64.841 | 170.076 | Infrastructure |
| Mix age building index | 0.812 | 0.764 | 0.172 | Infrastructure |
| Innovation spaces per 1000 residents | 0.0 | 0.023 | 0.319 | Infrastructure |
| Parks per 1000 residents | 0.584 | 1.148 | 2.34 | Urban Morphology |
| Squares per 1000 residents | 0.038 | 0.122 | 0.272 | Urban Morphology |
| Park land (acres) per 1000 residents | 7.774 | 155.055 | 1004.22 | Urban Morphology |
| Square land (acres) per 1000 residents | 0.008 | 0.056 | 0.144 | Urban Morphology |
| Car truck van to work pct | 0.886 | 0.807 | 0.214 | Urban Mobility |
| Public transportation to work pct | 0.01 | 0.087 | 0.167 | Urban Mobility |
| Walk bike to work pct | 0.025 | 0.048 | 0.083 | Urban Mobility |
| Worked from home pct | 0.036 | 0.046 | 0.051 | Urban Mobility |
| Worked outside state of residence pct | 0.01 | 0.039 | 0.077 | Urban Mobility |
| Bus stops per 1000 | 0.036 | 0.046 | 0.051 | Urban Mobility |
| Patents per 1000 residents (2016) | 0.0 | 58.393 | 1094.826 | Innovation Outcomes |
| SFR (2016) | 16.0 | 86.887 | 167.012 | Innovation Outcomes |

*Note: This table reports summary statistics for neighborhoods' elements zip-year observations from 2012 for New York State. There are 1856 such pairs in the dataset, as the inner join is used to keep only zip codes retrieved from the census dataset. The variables under the "Infrastructure" group and the rows under "Urban Mobility & Spatial Characteristic" present morphological characteristics, such as parks and squares, which have been retrieved more recently from OpenStreetMap, similar to the innovation spaces retrieved from the Google map; however, they have been used as the best approximation for 2012.*

**Table A3.** Assessing feature importance in innovation outcomes, having granted patents as the output, and using (i) all zip code data for Massachusetts and New York State, (ii) all zip codes for New York State, (iii) all zip codes for Massachusetts.

| Feature | Importance NY & MA | Importance NY | Importance MA |
|---|---|---|---|
| Firm R&D expenditure per 1000 residents | 0.17 | 0.256 | 0.114 |
| H1B Applications per 1000 residents | 0.126 | 0.133 | 0.471 |
| Business registrations per 1000 residents | 0.104 | 0.076 | 0.024 |
| 25 to 34 years pct | 0.102 | 0.087 | 0.001 |

| Feature | | | |
|---|---|---|---|
| Bus stops per 1000 residents | 0.096 | 0.117 | 0.007 |
| Cafes per 1000 residents | 0.087 | 0.09 | 0.088 |
| Population density | 0.071 | 0.023 | 0.002 |
| Graduate pct | 0.05 | 0.039 | 0.002 |
| Median income | 0.039 | 0.029 | 0.046 |
| Bachelor pct | 0.033 | 0.012 | 0.032 |
| Median age | 0.016 | 0.011 | 0.002 |
| Occupied housing units pct | 0.015 | 0.006 | 0.006 |
| Worked from home pct | 0.012 | 0.018 | 0.002 |
| Car truck van to work pct | 0.012 | 0.006 | 0.003 |
| Mix age building index | 0.011 | 0.008 | 0.004 |
| Park Land (Acres) per 1000 residents | 0.009 | 0.008 | 0.003 |
| Mean age of buildings | 0.006 | 0.007 | 0.006 |
| Native pct | 0.006 | 0.002 | 0.004 |
| Median home value | 0.005 | 0.003 | 0.002 |
| Walk bike to work pct | 0.004 | 0.014 | 0.006 |
| Parks per 1000 residents | 0.003 | 0.002 | 0.019 |
| White pct | 0.003 | 0.008 | 0.002 |
| Black pct | 0.003 | 0.007 | 0.007 |
| Public transportation to work pct | 0.003 | 0.002 | 0.001 |
| Asian pct | 0.003 | 0.004 | 0.002 |
| Poverty pct | 0.002 | 0.008 | 0.002 |
| Innovation spaces per 1000 residents | 0.002 | 0.001 | 0.115 |
| Unemployment rate | 0.002 | 0.004 | 0.001 |
| College pct | 0.002 | 0.002 | 0.002 |
| Scientific technical pct | 0.002 | 0.005 | 0.007 |
| Universities per 1000 residents | 0.001 | 0.005 | 0.002 |
| Worked outside state of residence pct | 0.001 | 0.002 | 0.002 |
| Schools per 1000 residents | 0.001 | 0.006 | 0.002 |
| Square Land (Acres) per 1000 residents | 0.0 | 0.0 | 0.006 |
| Squares per 1000 residents | 0.0 | 0.0 | 0.008 |

**Table A4.** Assessing feature importance in innovation outcomes, having start-up formation rate as the output and using (i) all zip code data for Massachusetts and New York State (ii) all zip codes for New York State (iii) all zip codes for Massachusetts.

| Feature | Importance NY & MA | Importance NY | Importance MA |
|---|---|---|---|
| Occupied Housing Units Pct | 0.513 | 0.579 | 0.544 |
| Median Home Value | 0.074 | 0.068 | 0.053 |
| Population Density | 0.071 | 0.025 | 0.008 |
| Public transportation to work pct | 0.057 | 0.035 | 0.007 |
| H1B application per 1000 residents | 0.037 | 0.038 | 0.072 |
| Business registrations per 1000 residents | 0.034 | 0.028 | 0.053 |
| Asian pct | 0.033 | 0.047 | 0.005 |
| Worked outside state of residence pct | 0.02 | 0.009 | 0.003 |
| 25 to 34 years pct | 0.014 | 0.011 | 0.015 |
| Black pct | 0.013 | 0.008 | 0.002 |
| Car truck van to work pct | 0.012 | 0.006 | 0.004 |
| Cafes per 1000 residents | 0.011 | 0.007 | 0.016 |
| Median age | 0.01 | 0.007 | 0.002 |
| Walk bike to work pct | 0.009 | 0.022 | 0.005 |
| R&D expenditure per 1000 residents | 0.009 | 0.01 | 0.077 |
| College pct | 0.007 | 0.039 | 0.011 |
| Bus stop per 1000 residents | 0.007 | 0.005 | 0.003 |
| Park Land (acres) per 1000 residents | 0.007 | 0.004 | 0.004 |
| Parks per 1000 residents | 0.006 | 0.005 | 0.003 |
| Poverty pct | 0.005 | 0.005 | 0.003 |
| White pct | 0.005 | 0.004 | 0.002 |
| Unemployment rate | 0.005 | 0.003 | 0.003 |
| Mean age of buildings | 0.005 | 0.004 | 0.003 |
| Mix age building index | 0.004 | 0.003 | 0.005 |
| Scientific technical pct | 0.004 | 0.005 | 0.024 |
| Median income | 0.004 | 0.003 | 0.007 |
| Graduate pct | 0.004 | 0.003 | 0.013 |
| Universities per 1000 residents | 0.004 | 0.004 | 0.002 |
| Schools per 1000 residents | 0.003 | 0.004 | 0.002 |
| Native pct | 0.003 | 0.002 | 0.003 |
| Worked from home pct | 0.003 | 0.003 | 0.004 |
| Bachelor pct | 0.003 | 0.003 | 0.014 |
| Innovation spaces total per 1000 residents | 0.002 | 0.002 | 0.005 |
| Square Land (Acres) per 1000 residents | 0.001 | 0.001 | 0.014 |
| Squares per 1000 residents | 0.001 | 0.001 | 0.019 |

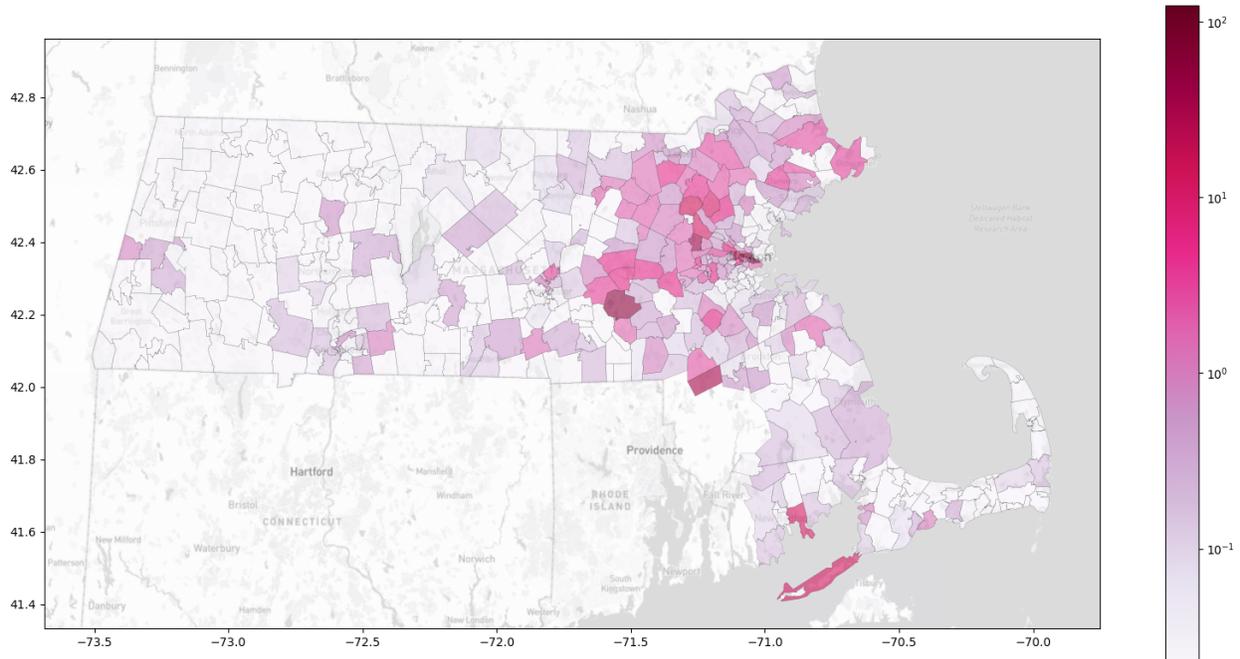

**Figure A1.** Map of the innovation output, the granted patent records per 1000 residents in 2016 across MA State's zip codes, and authors' elaboration.

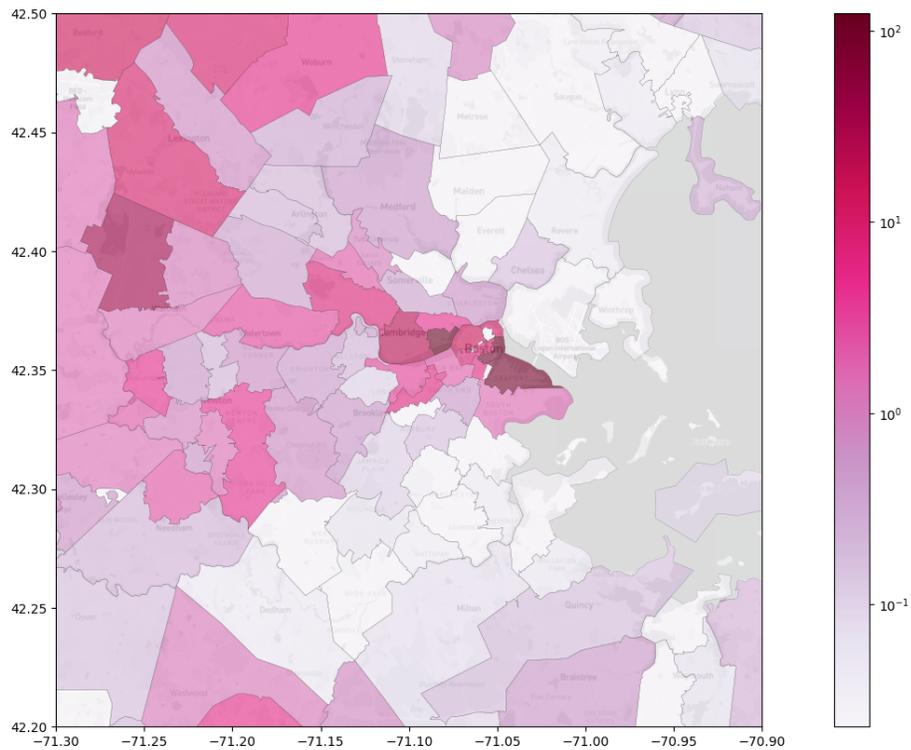

**Figure A2.** Map of the granted patent records per 1000 residents in 2016 across Boston City's zip codes, authors' elaboration.

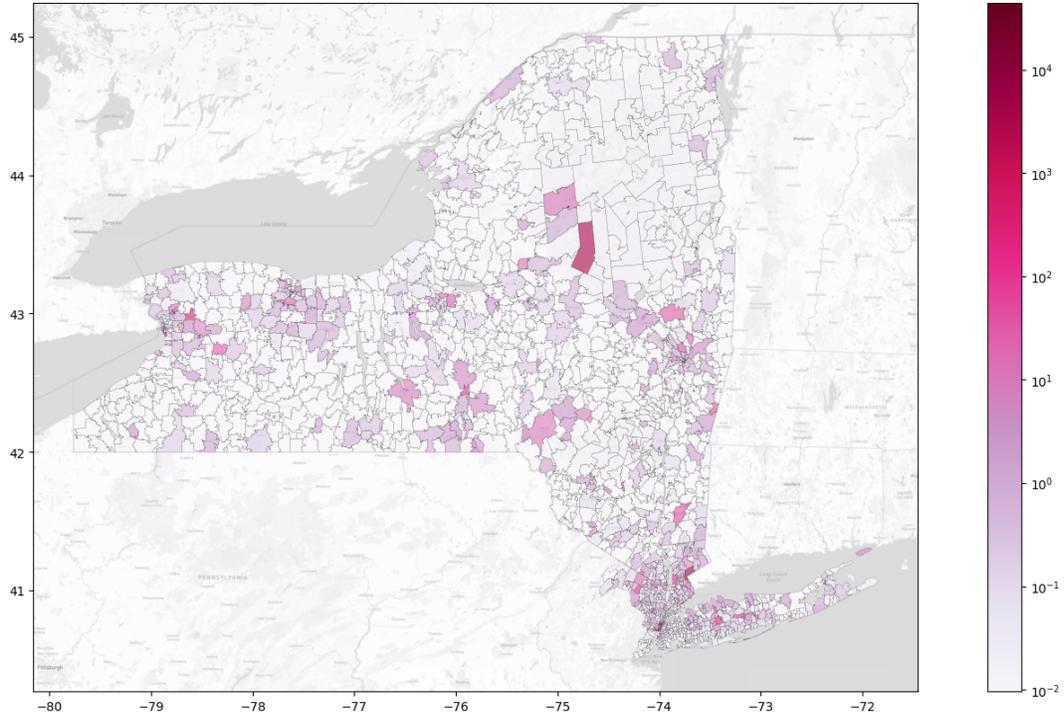

**Figure A3.** Map of the innovation output, the granted patent records in 2016 across New York State's zip codes, and authors' elaboration.

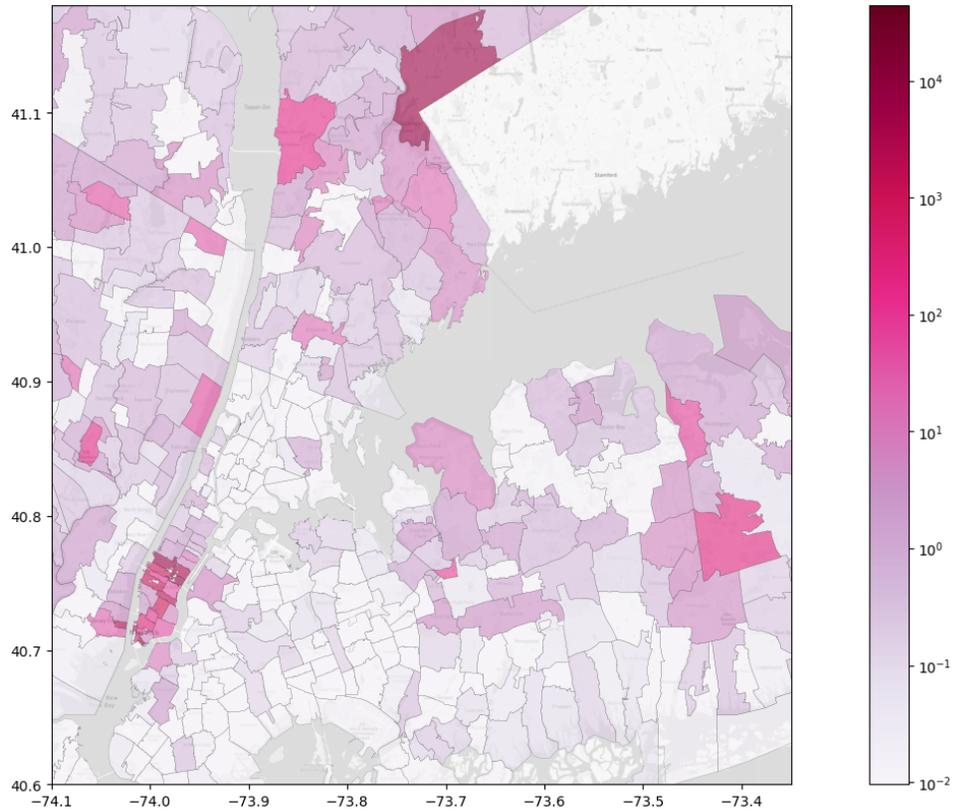

**Figure A4.** Map of the granted patent records in 2016 across the New York City metropolitan area's zip codes, authors' elaboration.

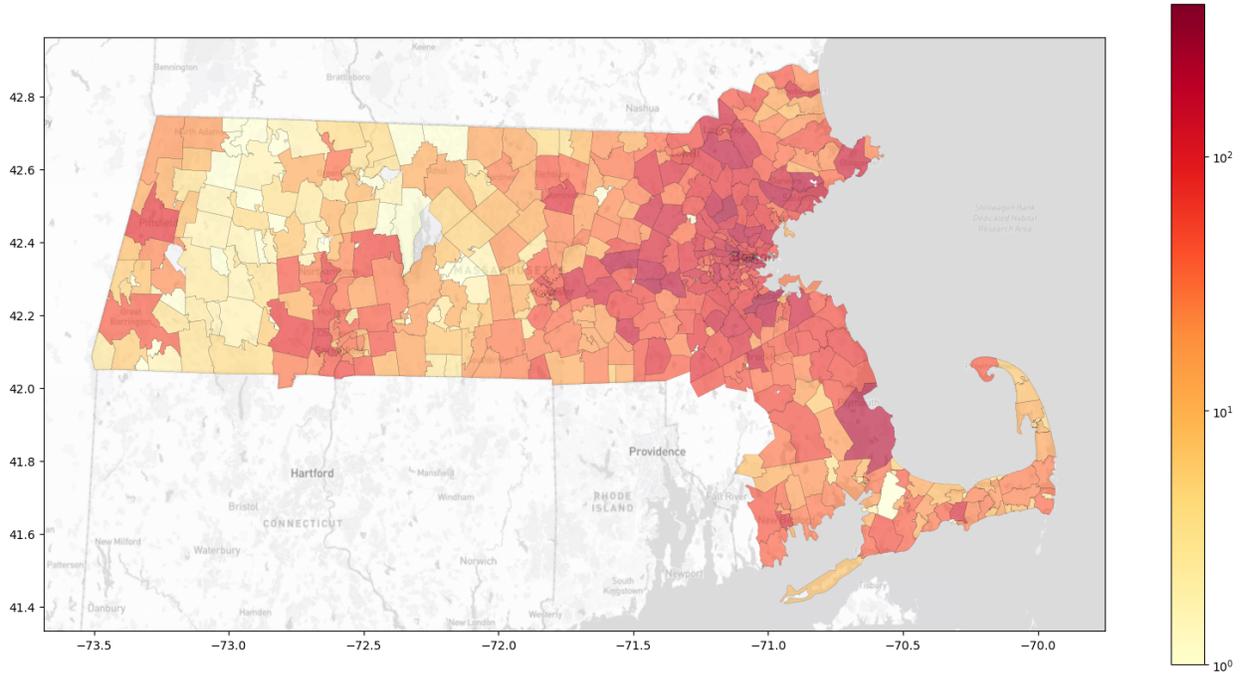

**Figure A5.** Map of the start-up formation rate in 2016 across MA State's zip codes, and authors' elaboration.

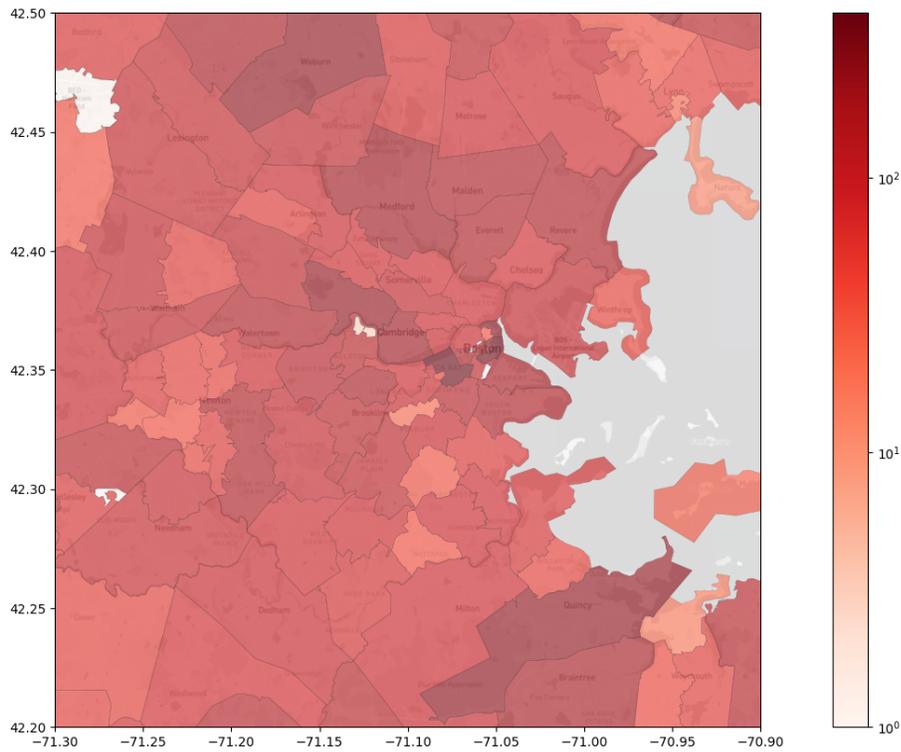

**Figure A6.** Map of the start-up formation rate in 2016 across Boston City's zip codes, authors' elaboration.

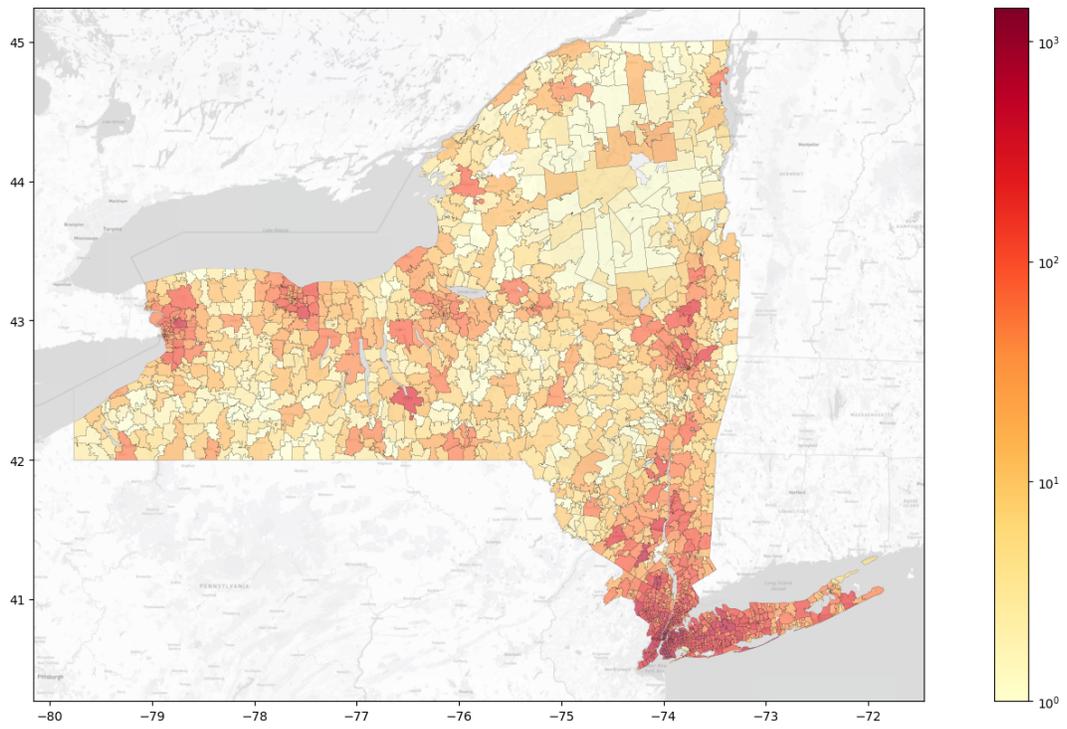

**Figure A7.** Map of the start-up formation rate in 2016 across New York State's zip codes, and authors' elaboration.

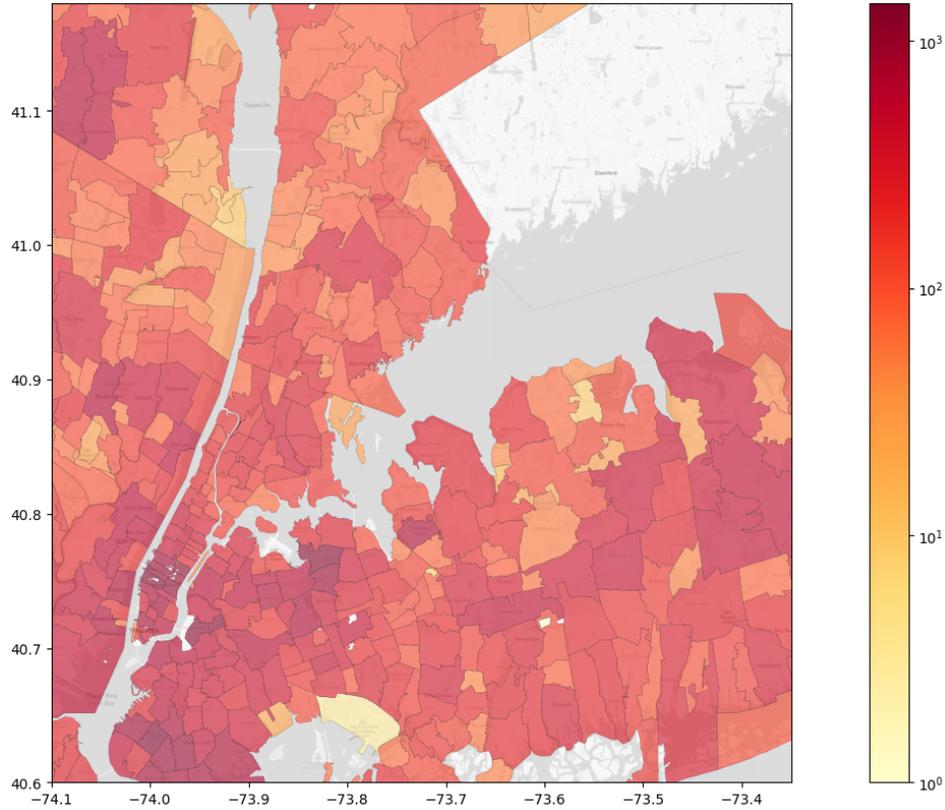

**Figure A8.** Map of the start-up formation rate records in 2016 across the New York City metropolitan area's zip codes, authors' elaboration.

**Author Contributions**

Conceptualization, E.O.; methodology, E.O.; Data collection, E.O., D.B.; investigation, E.O.; writing—original draft; preparation, E.O.; writing—review and editing, E.O., C.K.; project administration, E.O. All authors have read and agreed to the version of the manuscript.

**Acknowledgment**

Eleni thanks Prof. Jorge Guzman for the valuable comments during the conceptualization and data collection processes, which helped to formulate the project.